# Strategies for conformal REBCO windings

**J S Rogers[1], P M McIntyre[1,2], T Elliott[2], G D May[1] and C T Ratcliffe[1]**

[1]Texas A&M University
[2]Accelerator Technology Corp.

E-mail: jsr12e@tamu.edu

**Abstract**. **A high-field winding can be fabricated from a cable of non-insulated REBCO tapes, stacked face-face without twisting. If the cable is oriented within each turn of a winding so that the tape face is closely parallel to the magnetic field at its location, the supercurrent capacity of that cable is enhanced ~3x greater than in a transverse or twisting orientation. This concept for a conformal winding was presented in a previous paper pertinent to the body winding of a high-field dipole. Strategies are presented and simulated for how the same orientation can be sustained in the flared ends of a dipole.**

## 1. Introduction

REBCO tape has remarkable properties for use in superconducting magnets. It can operate with useful current density up to liquid nitrogen temperature, and it can produce very high magnetic field at temperatures of 20-40 K. The manufactured tape is ready to use as supplied, and does not require a final heat treatment after winding into a magnet winding – an important benefit compared to $Nb_3Sn$ and Bi-2212.

On the other hand, REBCO has certain undesirable properties that have limited its usefulness in high field magnets. REBCO is ruinously expensive, making it infeasible for use in many applications. Additionally, it is a highly anisotropic superconductor [1] with a critical current that is strongly dependent on the magnitude and direction of the magnetic field in which it is operating (Figure 2). The critical current is approximately 3x greater when the magnetic field at the tape is parallel to the tape face ($B_\parallel$, θ =90°) than when it is perpendicular ($B_\perp$, θ=0°).

In a previous paper a conformal winding method was presented in which a REBCO insert winding can be configured in a hybrid dipole so that the tapes within the winding are everywhere oriented parallel to the field at the tape, so that the winding can use the full capacity of the REBCO films. Figure 1 shows an example field design for such a dipole. The insert winding consists of a tape-stack cable in which 25 REBCO tapes are stacked face-to-face, and each turn of the cable is oriented closely parallel to the field at that location.

Clustering REBCO tapes has been used by several authors, including Roebel cable [2], CORC [3], cable-in-conduit [4], and stacked tapes [5]. In most cases REBCO tapes are stacked in a face-on cluster, multiple clusters are cabled around a solid copper core with a twist pitch so that each cluster spends equal length on the inside and outside of the cable in a winding. Yagotintsev *et al*. [6] review the measurements of AC loss and contact resistance in those several forms of clustered-tape conductors. But all of the present approaches to tape clustering involve twisting the clustter and so forego any possibility to sustain the $B_\parallel$ condition that would make it possible to use REBCO tape to its full potential. That is the goal of the conformal winding method.

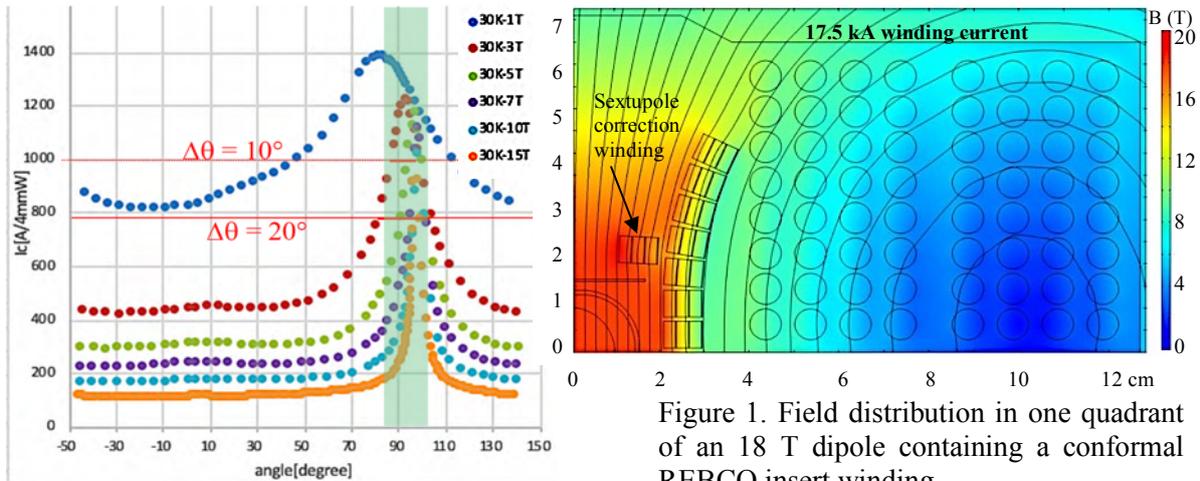

Figure 2. Dependence of $I_c$ on the angle between the tape face normal and the magnetic field, for various field strengths (30 K temperature) (from Ref. 1).

Figure 1. Field distribution in one quadrant of an 18 T dipole containing a conformal REBCO insert winding.

There are two challenges in making the conformal winding realizable:
- Can uniformly high critical current be sustained in the end windings of the dipole insert (where the magnetic field and the winding must both be flared) and in the leads, just as it is in the body?
- Can the tape-cluster cable support dynamic current-sharing within each cable turn so that inductive forces within the non-insulated cable do not concentrate its current in outer tapes and drive premature quench?

## 2. Optimization of conformal winding

The hybrid dipole shown in Figure 2 contains an insert winding made from rectangular REBCO tape-stack cable and an outsert winding made from $Nb_3Sn$ Cable-in-Conduit 'SuperCIC' [7].

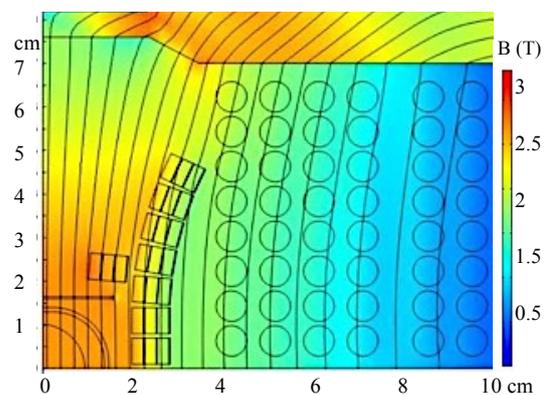

Figure 3. Field distribution at injection field.

summarizes the main parameters of both windings. The tape-stack cable contains 25 6-mm-wide REBCO tapes with 100 μm Cu clad to one side. The SuperCIC cable contains 17 0.85 mm-diameter wires of Hi-Lumi-class 108/127 Nb$_3$Sn/Cu wires, spiral-wrapped as a single layer around a thin-wall perforated center tube, then pulled as loose fit through a bronze sheath tube and drawn to compress the wires against the center tube to immobilize them.

**1.75 kA winding current**

Table 1. Main parameters of the 18 T hybrid dipole.

| | | |
|---|---|---|
| Bore field @ 4.2K short-sample | 18.3 | T |
| Cable current (windings in series) | 17.5 | kA |
| Aperture: horizontal, vertical | 40, 30 | mm |
| REBCO: #tapes/cable x #cables/shell x #shells | 25x 16 x 3 | |
| Nb$_3$Sn CIC: #wires/CIC x #cables/layer x #layers | 17x 16 x 8 | |
| B$_{max}$ in REBCO, CIC | 20.0, 11.8 | T |
| Sextupole @ full field, injection field | -5, -21 | units |

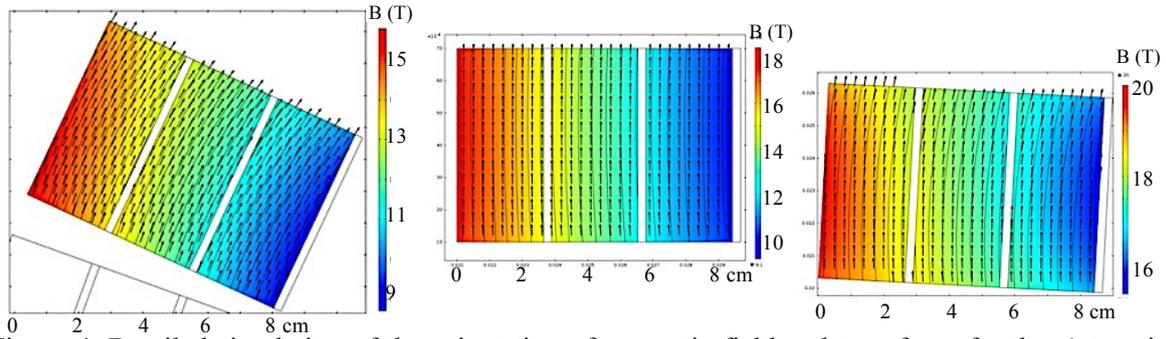

Figure 4. Detailed simulation of the orientation of magnetic field and tape faces for the a) top right; b) mid-plane; and c) sextupole 3-turn blocks of tape-stack cables in the REBCO insert winding.

The windings are operated in series, and the number of wires in each winding are chosen so that a cable current of 17.5 kA produces ~18 T bore field and corresponds approximately to critical current in each winding. The REBCO insert can operate at up to 20K, while the $Nb_3Sn$ winding operates at ~5K.

### 2.1. Conformal geometry at injection field

The field distribution in the winding region is significantly different at injection field and at collision field, due to progressive saturation in the flux return. Figure 3 shows the field distribution at injection field (1.75 kA winding current). There is a significant angle θ between the tape face and the field direction in several blocks, but the value of $|B|$ is low enough that the critical current $I_c(|B|, \theta)$ is sufficient that there is no risk of quenching providing that current-sharing is dynamically stable (the subject of the next section).

### 2.2. Body field simulations

The rectangular REBCO tape-stack cables are oriented along a curving contour to conform with the flaring behavior of the magnetic field in the region of the insert winding. The cylindrical CIC cables are oriented in a rectangular block-coil array in the outsert winding. There is also a 'sextupole correction' sub-winding of REBCO cables located inside and above the main REBCO insert, which is operated in series to correct sextupole in the bore field. Immediately above the bore tube is a steel flux plate that suppresses multipoles at injection field to suppress the effects of persistent currents and snap-back at injection field. The REBCO insert contributes $\sim 8\,T$ and the CIC winding $\sim 10\,T$ to the bore field.

Figure 4 shows a detailed cross-section of three particular 3-turn blocks within the REBCO insert winding: the top-right block, the mid-plane block, and the sextupole block. These blocks exhibit variously the highest field or the maximum deviation from the $B_\parallel$ orientation for some of the constituent tapes. The current capacity of the $n$th tape in each cluster can be estimated by extracting the local sheet critical current density $K_n(u)$ as a function of the location $u$ across the width of that tape, using the local values for $|B(x, y)|$ and θ, and adding the contributions for segments spanning the entire width $w$ of that tape: $I_{cn} = \int_0^w K_n(u)du$. The total cable current capacity is then obtained by adding the maximum current capacity of the 25 tapes in that cable segment. The maximum cable current for each cable is shown in the figure for each of the inner and outer cable turns.

### 2.3. End Field Simulations

An additional challenge is to design a winding strategy for the magnet ends that does not compromise the operational current capacity. In the 3D end regions of the dipole, fields flare and bend in a disorderly fashion raising concerns about the critical current of the cables. The flared ends of the tape-stack winding are formed using the method first pioneered by Willy Sampson (BNL) in the 1970s. Figure 5 shows a flared-end quadrupole containing a stacked-tape winding of dip-process $Nb_3Sn$ tape. Each tape-stack cable is twisted about its axis as it is flared vertically for form a catenary in which all tapes remain as a stack but no tape is bent in the hard direction. With this approach each tape face naturally follows closely the flaring of the magnetic field in the end.

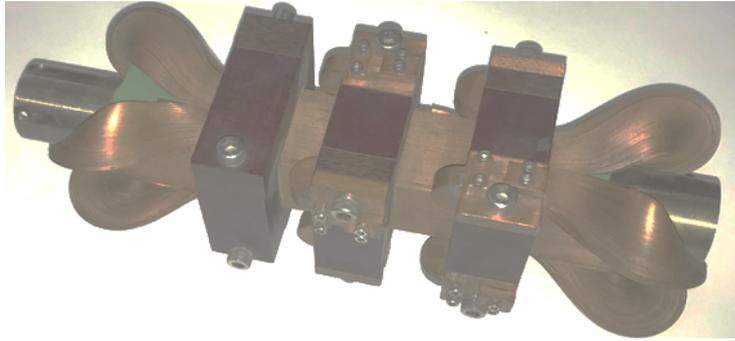
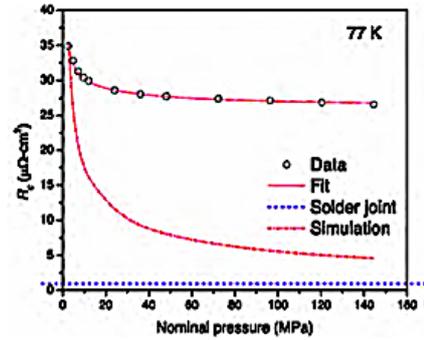

Figure 5. Example of flared ends in a tape-stack cable: Sampson's flared-end quadrupole using dip-coated Nb$_3$Sn superconducting tape.

Figure 6. Contact resistance between two copper-clad REBCO tapes as a function of compression.

This twisted-flare end has been modeled in a 3-D CAD design of the flared end, and the fields have been modeled using 3-D COMSOL. Figure 9 shows the results in three example cross sections through the end region: a) a y/z cross section through the vertical midplane; b) an x/y cross section at the transition from body to flared end; and c) an x/y cross section at a location 3 cm into the body from that transition. We are estimating $I_c(|B|,\theta)$ for each tape within each layer of each cross section using the data of Figure 2, and adding them to obtain the $I_c$ in each tape-stack cable at that location.

There is a compensating interplay that sustains $I_c$ with little or no degradation through the flared ends: the inner tapes are well-aligned to with the field direction but have the strongest value, while the outer tapes flare with significant angle with respect to the field but the field strength is low enough that the angular dependence of $I_c$ is also broadened. *The flared-end regions do not present a 'weak sister' limit to the $I_c$ of the cable in any turn of the winding.*

## 3. Dynamic current-sharing in the conformal winding

As the dipole is ramped to increase the bore field from its value $B_i$ at injection energy to $B_c$ at collision energy, the current within each tape-cluster cable is increased proportionately. As shown in Figure 8, the magnetic field at each cable produces a Lorentz force $\vec{F}$ that pushes current flowing within each tape-cluster cable away from the dipole bore. Thus, even if current were injected so that it was initially distributed equally among all tapes within a first turn of cable, the Lorentz force would re-distribute current to the outermost tapes within the winding. It would therefore seem that as coil current were increased the outermost tape would quench when the overall cable current was still only a fraction of the desired current!

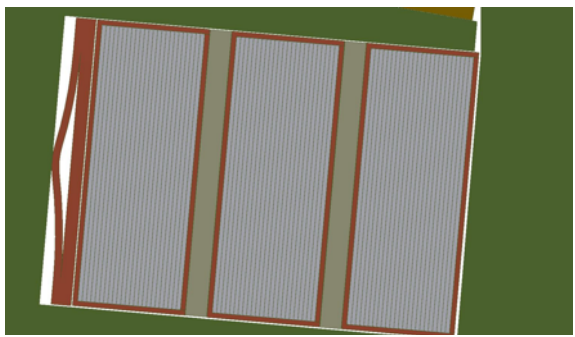
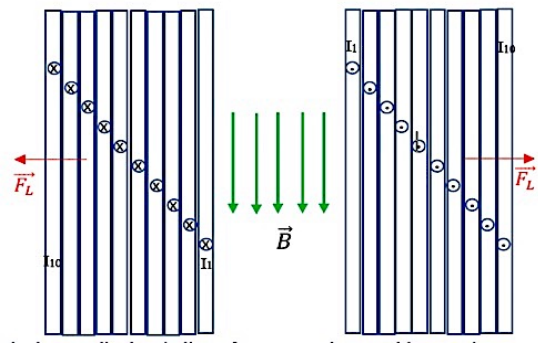

Figure 7. Detail showing a block of 3 tape-stack tapes, each containing 25 tapes, with a laminar spring to provide uniform ~1 MPa compression to all cables.

Figure 8. Schematic illustration of current-sharing dynamics in a conformal winding of tape-stack REBCO cable.

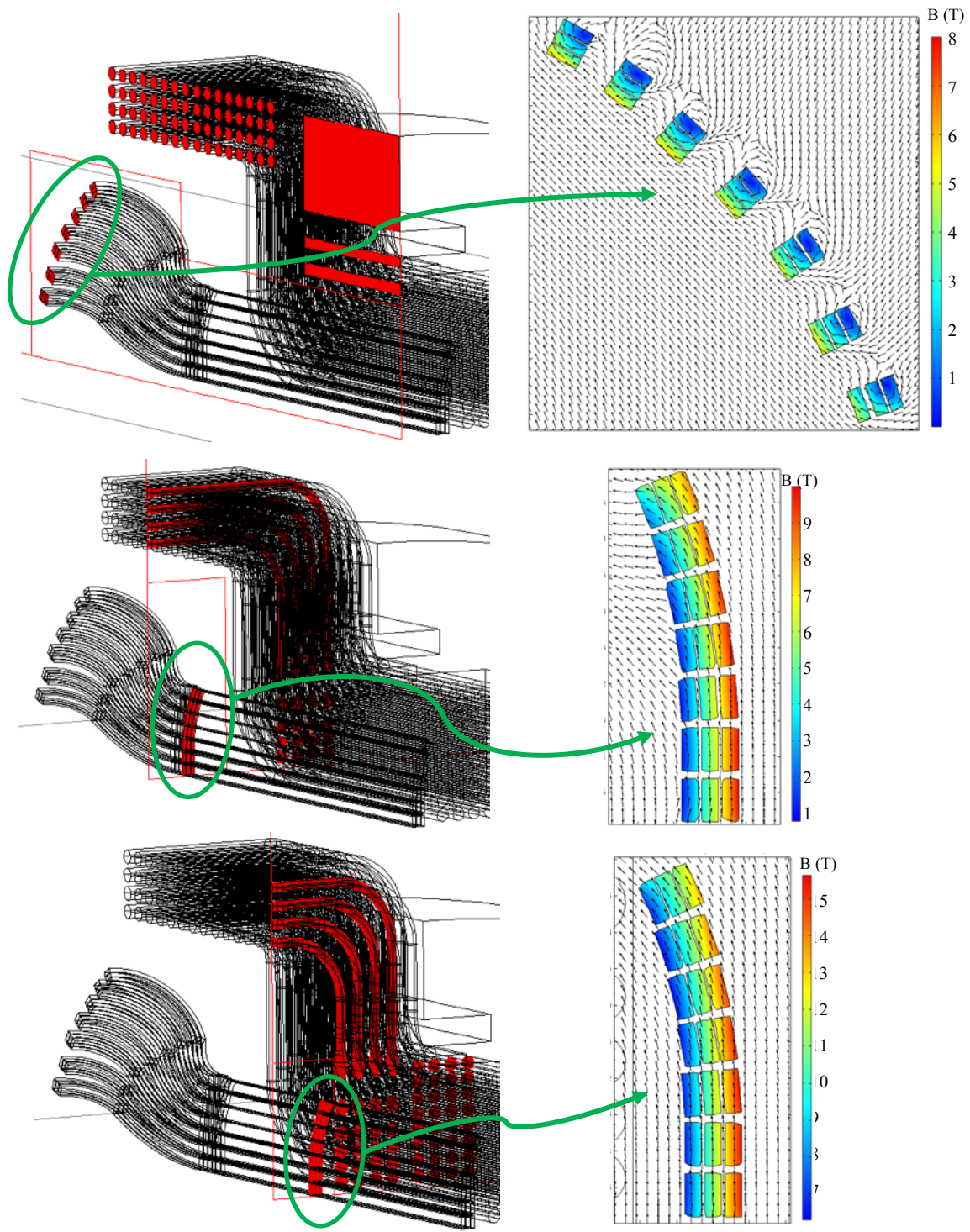

Figure 9. Three sections through the REBCO end winding of the hybrid dipole: a) section through the vertical midplane; b) x/y cross section at the transition from body to end winding; c) x/y cross-section 3 cm into the body winding.

But REBCO can operate at 30 K, where the heat capacity of the tape ($\sim T^3$), and the conduction to remove heat $\sim(T_{hot}-T)$ are both much greater than at liquid helium temperature, so we conjecture that it may be possible to operate a cable of stacked non-insulated tapes *without transposition* and rely upon the 'soft' approach to quench in each tape layer to force re-distribution of current to neighboring tapes within the cable as the cable current is further increased. This strategy has been used to good effect in 'no-insulator' (*NI*) pancake windings for high-field solenoids [8], but never in a dipole. We now analyze the dynamics of quasi-equilibrium current-sharing among the tapes within a non-transposed cable in the dipole of Figure 1, using the modelling approach of Noguchi [9].

The REBCO layer within each exhibits a retarding electric field E that is current dependent:

$$E_z = E_0 (I/I_c)^n \tag{1}$$

where $E_0 = 10^{-6}$ V/m is the quench criterion, $I_c(|B|, \theta, T)$ is the critical current for the conditions $(B, \theta, T)$ for that tape, and $n \sim 30$ is the index that characterizes the power-law dependence of the superconductor-normal transition for REBCO.

The dynamics is analogous to the Hall effect, in which the superconducting transport is acted upon by the transverse Lorentz force, by a transverse electric field produced by the potential difference between neighboring tapes when they carry different currents, and by a contact resistivity $R_c$ between adjacent tapes through which current is displaced by Lorentz forces.

Following Ref. 9, the time dependent distribution of current in a tape-stack cable can be estimated in a simple model in which the full length of one half-turn of the tape-stack cable is treated as a series-parallel L/R network. Each tape within a half-turn of one cable has a self-inductance per unit length

$$\tilde{L} = \frac{\mu_0 x}{w g} = 4 \times 10^{-4} H/m \tag{2}$$

and a power-law series resistance/length

$$\tilde{R}_s = \frac{E}{I} = \frac{E_0}{I_0}\left(\frac{I}{I_0}\right)^{n-1} = (0.7 n\Omega/m)\left(\frac{I}{I_0}\right)^{n-1} \tag{3}$$

where *w*=6 mm is the tape width, *g*=10 cm is the vertical gap in the steel flux return, and *x*~10 cm is the horizontal width of the tape loop.

Lu *et al*. [10] measured the dependence of the contact surface resistance $R_c$ upon the compression among the tapes in the stack, shown in Figure 6. As shown in Figure 7, each turn of tape-cluster cable in the conformal winding is supported by a laminar spring that provides ~1 MPa uniform compression all tapes of the tape-cluster, corresponding to $R_c$ ~35 µΩ-cm². The parallel resistance $R_p$ by which a tape segment shares current with its neighbors is obtained by dividing $R_c$ by the segment area (length x width):

$$R_p = \frac{R_c}{\mathcal{L} w} = \frac{(0.6 \mu\Omega \cdot m)}{\mathcal{L}} \tag{4}$$

From these quantities, we can extract two results that characterize the scale of current-sharing. First, the scale length $\lambda$ over which this homogenization operates is the winding length for which $R_p \sim R_s \lambda$:

$$\lambda = \sqrt{\frac{R_c}{w \tilde{R}_s}} = 29 m \left(\frac{I}{I_c}\right)^{-11.5} \tag{5}$$

$\lambda$ is much longer than any reasonable winding length, so the current distribution would relax uniformly along the winding.

Second, we can estimate the time constant with which a difference in current between successive tapes in a tape-stack cable relaxes to an equilibrium governed by the Lorentz force and the 2-D distribution of resistance within a tape-stack cable. The change in inductance *dL* along one winding length $\mathcal{L}$ between one tape and the next is

$$dL = \frac{\mu_0}{w g}\mathcal{L}; \qquad dL[\mu H] = 0.4 \mathcal{L}[m] \tag{6}$$

So the time constant for relaxation between tapes is $\tau = \frac{dL}{R_p}; \qquad \tau[s] \sim (1 \text{ second}) \mathcal{L}[m]^2 \tag{7}$

In a conformal winding for a collider dipole, the current re-distributes rapidly enough that no tape should reach $I_c$ prematurely.

From this simple model, we predict that, as coil current is increased from zero, current would accumulate in the outermost tape of each tape-stack cable until the coil current approaches a limit $I_0 \sim 0.8$ *NI*$_c$(*B,T,θ*) for that tape. Then as coil current is further increased, current would share to the neighboring tape until the coil current reached *2 I*$_c$ in the two tapes. Then as coil current is further increased, current would share to the 3$^{rd}$ tape, etc., until finally current would become ~homogeneous throughout the cable as the current approached an ultimate limit of ~*NI*$_0$.

There is interesting physics in the current-sharing among *NI* REBCO tapes that are face-aligned with $\vec{B}$. It is our hope that the above simple model will stimulate others to develop a multi-scale model that connects the normal transition at nanoscale to the redistribution of currents and forces among the tapes.

## 4. Field homogeneity for collider requirements

Field homogeneity is of particular concern for the dipole magnets of an accelerator or collider. The sextupole harmonic can be selectively canceled by placement of one correction turn in the winding, at the location shown in Figure 1. The particular example magnetic design shown has been optimized to produce nearly pure dipole field over a dynamic range of field 0.2-4 T, in which the amplitudes $b_n$ are all <$10^{-4}$ over that range.

Current-sharing poses a further challenge for field homogeneity, however. At injection field, the current in each tape cluster is located mainly in the outermost tape; at collision field, the current is ~equally shared, so the 'current position' for that cable turn is shifted inwards by half the cluster width.

Conventionally the $n^{th}$ multipole of a dipole field distribution is defined at each location $\vec{r} = r(\hat{x}cos\theta + \hat{y}sin\theta)$ by the expansion $B_x + iB_y = \sum b_n e^{-ni\theta} \left(\frac{r}{R}\right)^{n-1}$ (8)

The multipoles have been evaluated for the magnetic design of Figure 1 for the maximum field ( limiting cases. The difference in the calculated multipoles is $\Delta b_n$ <$0.5 \times 10^{-4}$ for all multipoles! This remarkable result is a consequence of the conformal design strategy: because each tape cluster is oriented so its face is closely parallel to $\vec{B}$, the field distribution is insensitive to the horizontal position of the 'current center position' of that cluster.

Prestemon *et al*. [11] modeled current transfer in a stacked-tape cable and found that a region of cable with coupling resistance $R \sim 10\ n\Omega$ is effective in equalizing current among its tapes. The laminar spring sustains 1 MPa compression, so the characteristic length of cable for stability is

$$L \sim \frac{35 \mu\Omega \cdot cm^2}{10 n\Omega \cdot 6mm} = 60\text{m} \quad (9)$$

This is conveniently the approximate length for one turn of a collider dipole so ramping current should stabilize turn-by-turn.

## 5. Conclusion

A strategy is presented by which a cable containing multiple REBCO tapes may be configured in a magnet winding in such a way that the favorable $B_∥$ orientation is sustained everywhere in the winding. So that the REBCO tapes could perform to their maximum potential. A method for dynamic current sharing is discussed, in which current would naturally re-distribute among the tapes of each cable turn as windng current is increased without inducing premature quench.